\documentclass[slac_one]{revtex4}
\usepackage{graphicx}
\usepackage{fancyhdr}
\newcommand{\abs}[1]{\ensuremath{\left|#1\right|}}
\newcommand{\ks}  {\ensuremath{K_{S}}}
\newcommand{\kl}  {\ensuremath{K_{L}}}
\newcommand{\kpm} {\ensuremath{K^{\pm}}}
\newcommand{\kp} {\ensuremath{K^{+}}}
\newcommand{\km} {\ensuremath{K^{-}}}
\newcommand{\pipm} {\ensuremath{\pi^{\pm}}}
\newcommand{\vus}{\ensuremath{V_{us}}}
\newcommand{\vud}{\ensuremath{V_{ud}}}
\newcommand{\vub} {\ensuremath{V_\mathrm{ub}}}
\newcommand{\fVus}{\ensuremath{f_+(0)\,V_{us}}}
\newcommand{\fzero}  {\ensuremath{f_+(0)}}
\newcommand{\fzerokpi}  {\ensuremath{f_+^{K^0\pi^-}(0)}}
\newcommand{\BR} {\ensuremath{\mathrm{ BR}}}
\newcommand{\kzero} {\ensuremath{K^{0}}}
\newcommand{\kltre}  {\ensuremath{K_{l3}}}
\newcommand{\ketre}  {\ensuremath{K_{e3}}}
\newcommand{\kletre}  {\ensuremath{K_{Le3}}}

\newcommand{\ksetre}  {\ensuremath{K_{Se3}}}
\newcommand{\kmutre}  {\ensuremath{K_{\mu3}}}
\newcommand{\klmutre}  {\ensuremath{K_{L\mu3}}}

\newcommand{\kpmetre}   {\ensuremath{K^{\pm}_{e3}}}
\newcommand{\kpmmutre}  {\ensuremath{K^{\pm}_{\mu3}}}

\newcommand{\pippimpio}  {\ensuremath{\pi^+\pi^-\pi^0}}
\newcommand{\pippim}  {\ensuremath{\pi^+\pi^-}}
\newcommand{\duepio}  {\ensuremath{2\pi^0}}
\newcommand{\trepio}  {\ensuremath{3\pi^0}}
\newcommand{\mudue}  {\ensuremath{\mu^{\pm}\nu}}

\newcommand{\taus}  {\ensuremath{ \tau_S}}
\newcommand{\taul}  {\ensuremath{\tau_L}}
\newcommand{\taupm}  {\ensuremath{\tau_{\pm}}}
\newcommand{\lv}       {\ensuremath{\lambda_+}}
\newcommand{\lvp}       {\ensuremath{\lambda^{\prime}_+}}
\newcommand{\lvpp}       {\ensuremath{\lambda^{\prime\prime}_+}}
\newcommand{\ls}       {\ensuremath{\lambda_0}}
\newcommand{\lsp}       {\ensuremath{\lambda^{\prime}_0}}
\newcommand{\lspp}       {\ensuremath{\lambda^{\prime\prime}_0}}

\let\kpppg=\kppipigall

\def\ff{$\phi-$factory}  \def\DAF{DA\char8NE}
  
\def\pt#1,#2,{\ifm{#1\x10^{#2}}}

\begin{document}

\title{Unitarity and universality with kaon physics at KLOE} 

%

\author{Erika De Lucia for the KLOE Collaboration \footnote{ 
F.~Ambrosino,
A.~Antonelli, 
M.~Antonelli, 
F.~Archilli,
C.~Bacci,
P.~Beltrame,
G.~Bencivenni, 
S.~Bertolucci, 
C.~Bini, 
C.~Bloise, 
S.~Bocchetta, 
F.~Bossi,
P.~Branchini,
P.~Campana, 
G.~Capon, 
T.~Capussela,
F.~Ceradini,
S.~Chi,
G.~Chiefari, 
P.~Ciambrone,
E.~De~Lucia,
A.~De~Santis, 
P.~De~Simone, 
G.~De~Zorzi,
A.~Denig,
A.~Di~Domenico,
C.~Di~Donato,
B.~Di~Micco,
A.~Doria,
M.~Dreucci,
G.~Felici, 
A.~Ferrari,
M.~L.~Ferrer, 
S.~Fiore,
C.~Forti,       
P.~Franzini,
C.~Gatti,      
P.~Gauzzi,
S.~Giovannella,
E.~Gorini, 
E.~Graziani,
W.~Kluge,
V.~Kulikov,
F.~Lacava, 
G.~Lanfranchi, 
J.~Lee-Franzini,
D.~Leone,
M.~Martini,
P.~Massarotti,
W.~Mei,
S.~Meola,
S.~Miscetti, 
M.~Moulson,
S.~M\"uller,
F.~Murtas, 
M.~Napolitano,
F.~Nguyen,
M.~Palutan,          
E.~Pasqualucci,
A.~Passeri,  
V.~Patera,
F.~Perfetto,
M.~Primavera,
P.~Santangelo,
G.~Saracino,
B.~Sciascia,
A.~Sciubba,
A.~Sibidanov,
T.~Spadaro,
M.~Testa,
L.~Tortora, 
P.~Valente,
G.~Venanzoni,
R.~Versaci,
G.~Xu.
}
}

\affiliation{Laboratori Nazionali di Frascati dell'INFN, Roma, Italia.}

\begin{abstract}
All relevant inputs for the extraction of the CKM matrix element
 \vus\ from \kl, \ks\ and \kpm\ decays have been measured at KLOE.
From a global fit using only KLOE results, but \ks\ lifetime, a value of
$|\vus|\fzero = 0.2157 \pm 0.0006$ 
is obtained, where \fzero\ is the form factor parametrizing the hadronic
matrix element evaluated at  zero momentum transfer. 
Comparison of the values of \fVus\ for $K_{e3}$ and $K_{\mu3}$ modes provides a test of lepton
universality at 0.8\% level of accuracy.
The value $\vus/\vud=0.2323(15)$ has been obtained
from the ratio
$\Gamma(K\to\mu \nu)/\Gamma(\pi\to\mu \nu)$
using KLOE 
measurement of $\BR(\kpm \rightarrow \mudue)$  
and lattice calculation of 
the ratio of decay constants $f_K/f_{\pi}$.
These results, together with $\vud = 0.97418(26)$, are compatible at
0.6$\sigma$ level with CKM matrix unitarity. 
The universality of lepton and quark weak couplings can be tested and
constraints on new physics extensions of the Standard Model can be set
using these very precise measurements from kaon decays.
\end{abstract}

\maketitle

\thispagestyle{fancy}


\section{INTRODUCTION} 
Precise measurements of semileptonic kaon decay rates 
provide the measurement of the \vus\ element of the CKM mixing matrix
and information about lepton universality. 
Helicity-suppressed leptonic kaon decays provide 
an independent measurement of $\abs{\vus}^2/\abs{\vud}^2$, through  
the ratio  $\Gamma(K\to\mu \nu)/\Gamma(\pi\to\mu \nu)$.
These measurements, together with the
result of $|\vud|$ from nuclear $\beta$ transitions,
provide the most precise test of the unitarity of the CKM mixing matrix through the relation 
$|\vud|^2 + |\vus|^2 + |\vub|^2=1$. 
Unitarity can also be interpreted as a test of 
the universality of lepton and quark weak couplings, testing the relation 
$G^2_F = G^2_{CKM}=G^2_F(\vud^2 +\vus^2+\vub^2)$, with $G_F$ the Fermi coupling constant from the muon decay.
Bounds on new physics extensions of the standard model can be set
using the ratio of the \vus\ values obtained from helicity-suppressed
$K_{\ell2}$ decays and helicity-allowed $K_{\ell3}$ decays.
The kaon semileptonic decay rate is given by:  
\begin{equation}
\Gamma(\kltre) = \frac{C_K^2 G_F^2 M_K^5}{192 \pi^3} S_{EW} |\vus|^2 |\fzero|^2 
I_{K,l}(\lambda)(1+2\Delta_K^{SU(2)}+2\Delta_{K,l}^{EM})
\label{eq:gammakl3}
\end{equation}
where $K = \kzero, \kpm$, $l = e, \mu$ and $C_K$ is a Clebsch-Gordan coefficient, equal to $1/2$ and $1$ for \kpm\ 
and \kzero, respectively. The decay width $\Gamma(\kltre)$ is experimentally determined 
by measuring the kaon lifetime and the semileptonic BRs totally inclusive of radiation.
The theoretical inputs are: the universal 
short-distance electroweak correction $S_{EW} = 1.0232$, 
the $SU(2)$-breaking $\Delta_K^{SU(2)}$ and the long-distance electromagnetic
corrections $\Delta_{K,l}^{EM}$, which depend on the kaon charge and on the lepton flavor, and the form factor $\fzero\equiv\fzerokpi$ parametrizing the hadronic matrix element of 
the $K \rightarrow \pi$ transition, evaluated at zero momentum transfer and for neutral kaons. 
The form factor dependence on the momentum transfer can be described by one or more slope parameters $\lambda$, 
measured from the decay spectra, and enters in the phase space integral $I_{K,l}(\lambda)$.

The KLOE experiment at the Frascati \ff\ \DAF\ can measure all the relevant inputs
to extract \vus\ from \kltre\ decay rates of both charged and neutral kaons: \BR s, lifetimes and form factors.
In section~\ref{sec:obs} an overview of the measurements for \kl, \ks\ and \kpm\ 
will be presented, followed by the extraction of \vus\ in
section~\ref{sec:vusf} and bounds on new physics beyond the standard model in section~\ref{sec:np}. 
\section{BRANCHING RATIOS, LIFETIMES AND FORM FACTOR SLOPES}
\label{sec:obs}
\subsection{\kl\ decays}
\label{subsec:kl}
The absolute BRs for the four main \kl\ decay channels have been measured
from a sample of $1.3\times 10^7$ $\phi \rightarrow \ks\kl$ events tagged by $\ks \rightarrow \pippim$ decays.
 The results depend on the \kl\ lifetime $\taul$ through the geometrical acceptance of the apparatus.
Using as reference value $\taul = 51.54$~ns we get  
$\BR(\kletre) = 0.4049(21)$, $\BR(\klmutre) = 0.2726(16)$,
$\BR(\trepio) = 0.2018(24)$, and $\BR(\pippimpio) = 0.1276(15)$~\cite{kloebrl}.
An independent measurement, $\taul=50.92(30)$~ns~\cite{kloetaul}, 
is obtained from the fit of the proper decay time distribution
for $\kl \rightarrow \trepio$ events.
\kl\ tagged decays have been used for the measurement of the form factors
for both \ketre\ and \kmutre\ decays.
For \ketre\ decays the fit of the $t/m_{\pi}^{2}$ distribution, with $t$ the squared momentum transfer,
is sensitive to both linear and quadratic terms of the power expansion of the form factor
and agreement has been found between the results obtained with quadratic and pole parametrization. 
With the quadratic parametrization we get: $\lvp=25.5(1.8)\times10^{-3}$ and $\lvpp=1.4(0.8)\times10^{-3}$~\cite{kloe_ffe3}.
Using \kmutre\ decays,
the $\ls$ slope of the scalar form factor is obtained fitting the
distribution of the neutrino energy $E_{\nu}$,due to the
difficult $\pi/\mu$ identification at low energies, while $\lvp$ and $\lvpp$   
are measured from the combined fit with \ketre\ decays. The results are:
$\lvp=25.6(1.7)\times10^{-3}$, $\lvpp=1.5(0.8)\times10^{-3}$ and 
$\ls=15.5(2.2)\times10^{-3}$~\cite{kloe_ffm3}. 
The fit to data distributions is only sensititive to the linear term of the
power expansion of the scalar form factor, due to the -99.6\% correlation
between the coefficients of the linear \lsp\  and quadratic terms \lspp. 
However the linear parametrization over-estimates the value of \lsp by
about 20\%. The solution is to use  the dispersive representation in which vector and scalar form factors are
described by two parameters only\cite{disper}:$\lv$ and $\ls$. The results are $\lv=25.7(0.6)\times10^{-3}$ and
$\ls=14.0(2.1)\times10^{-3}$~\cite{kloe_ffm3} and have been used in
the evaluation of \vus. From a  preliminary update based on a larger \kmutre\ data
sample, averaged with the published values, we get: $\lv=26.0(0.5)\times10^{-3}$ and $\ls=15.1(1.4)\times10^{-3}$.
\subsection{\ks\ decays}
\label{subsec:ks}
A \ff\ provides the unique opportunity of having a pure \ks$-$beam.
Therefore, using a sample of $1.2\times 10^8$ $\phi\rightarrow\ks\kl$ events in which the \kl\ is 
identified by its interaction in the calorimeter, we have measured 
$\Gamma(\ksetre)/\Gamma(\pippim) = 10.19(13) \times 10^{-4}$~\cite{kloekse3} and 
$\Gamma(\pippim)/\Gamma(\duepio) = 2.2549(54)$~\cite{kloeks2pi}.
These two ratios completely determine the value of \ks\ main BRs 
and allow us to measure $\BR(\ksetre) = 7.046(91)\times 10^{-4}$ to be used in the \vus\ extraction.
The value of \ks\ lifetime used in the present determination of $|\vus|\fzero$ is 
$\taus = 0.08958(5)$~ns, from PDG fit to $CP$ parameters~\cite{PDG06}.
\subsection{\kpm\ decays}
\label{subsec:kpm}
The values of BR($\kpmetre$) and BR($\kpmmutre$) have been determined separately 
for $K^+$ and $K^-$ decays tagged by \kp\ and \km\  so to have four
independent measurements for each BR, minimizing the systematic contribution from the tagging procedure. The background rejection uses kinematics 
and the signal count is extracted from a constrained likelihood fit 
to the distribution of the squared lepton mass evaluated by time-of-flight measurements.
Using $\taupm = 12.385(25)$~ns~\cite{PDG06} to account 
for acceptance dependence on kaon lifetime ($\taupm$), we obtain $\BR(\kpmetre) = 0.04965(53)$, and 
 $\BR(\kpmmutre) = 0.03233(39)$~\cite{kpmkl3}. 
The world average accuracy on \kpm\ lifetime \taupm\  is
0.2$\%$~\cite{PDG06} however the measurements\'\ spread signals a poor consistency  
between the results. 
The KLOE experiment has measured \taupm\ with $\kpm \rightarrow \mudue$
tagged kaons and fitting the distribution of the kaon 
proper decay time $t^{\ast}$ evaluated with two independent methods: the
kaon decay length and the kaon decay time from the time of 
flight of the photons from the $\pi^0$ in the final state.
The first method gives $\taupm = 12.364(31)(31)$~ns, the second $\taupm
= 12.337(30)(20)$~ns and the average is $\taupm = 12.347(30)$~ns\cite{tau_kpm}, in perfect
agreement with PDG value. The absolute branching ratio of the \kpppg\ decay has been measured 
using $\sim$20 million tagged $K^{+}$ 
mesons. Signal counts are obtained from the fit of the distribution
of the momentum of the charged decay particle in the kaon rest frame. 
The result, inclusive of final-state radiation, is
\BR(\kpppg)=0.2065(9)~\cite{kpi2}, 1.3\% ($\sim$2$\sigma$) lower than the PDG fit~\cite{PDG06}.
\begin{table}[!h]
\begin{center}
\begin{tabular}{|c|c|c|c|c|c|c|} 
\hline
 $\BR(K^+_{\mu2})$  & $\BR(K^+_{\pi2})$ &
 $\BR(\pi^\pm\pi^+\pi^-))$ & $\BR(K^\pm_{e3})$ & $\BR(K^\pm_{\mu3})$ &
 $\BR(\pi^{\pm}\pi^0\pi^0)$ &  \taupm (ns) \\ \hline
 0.6376(12) & 0.2071(9) & 0.0553(9) & 0.0498(5) & 0.0324(4) &
 0.01765(25) & 12.344(29) \\ \hline
\end{tabular}
\caption{\label{tab:kpmfit} Results of the fit to \kpm\ BRs and lifetime.}
\end{center}
\vspace{-0.6cm}
\end{table}
We then fit the six largest \kpm\ BRs and the lifetime $\tau$ 
using our measurements of \taupm ~\cite{tau_kpm}, $\BR{K^+_{\pi2}}$\cite{kpi2}
$\BR{K^+_{\mu2}}$~\cite{kmu2}, $\BR{K^\pm_{\rm l3}}$~\cite{kpmkl3} and
$\BR{K^{\pm}\to\pi^{\pm}\pi^0\pi^0}$~\cite{plbtauprime}, with their
dependence on \taupm, together with 
$\BR(\kpm\to\pi^\pm\pi^+\pi^-)$ from the
PDG04 average~\cite{pdg04}, 
with the sum of the BRs constrained to unity. 
The fit result, listed in Table~\ref{tab:kpmfit}, with $\chi^2/{\rm ndf}=0.59/1$ (CL=44\%) shows very good
consistency~\cite{kpi2}.
\section{EXTRACTION OF $|\vus|\fzero$ AND $|\vus|$}
\label{sec:vusf}
To extract $|\vus| \fzero$ we need the $SU(2)$-breaking~\cite{su2breaking} and the 
long distance $EM$ corrections to the full inclusive decay rate~\cite{su2breaking,moussallam},
evaluated for the first time for $\kmutre$ channels of both neutral and charged kaons~\cite{neufeld}.
The values of $|\vus|\fzero$ measured for 
$K_Le3$, $K_L\mu3$, $K_Se3$, $K^{\pm}e3$, and $K^{\pm}\mu3$ decay modes are shown in Table~\ref{tab:vusf}.
The five decay modes agree well within the quoted errors and average to $|\vus|\fzero = 0.2157\pm0.0006$, with 
$\chi^2/ndf =7.0/4$ (Prob=13\%), to be compared with the world average $|\vus|\fzero = 0.2166 \pm 0.0005$~\cite{allvus}.
\begin{table}[!h]
\begin{center}
\begin{tabular}{|c|c|c|c|c|c|} 
\hline
 Mode          & $K_Le3$  & $K_L\mu3$ & $K_Se3$ & $K^{\pm}e3$ &
 $K^{\pm}\mu3$ \\ \hline
 $|\vus|\fzero$ & 0.2155(7) & 0.2167(9) & 0.2153(14) & 0.2152(13) &
 0.2132(15) \\ \hline
\end{tabular}
\caption{\label{tab:vusf}Values of $|\vus| \fzero$ extracted from \kltre\ decay rates.}
\end{center}
\vspace{-0.4cm}
\end{table}

\begin{figure}[!h]
\includegraphics[width=7.cm]{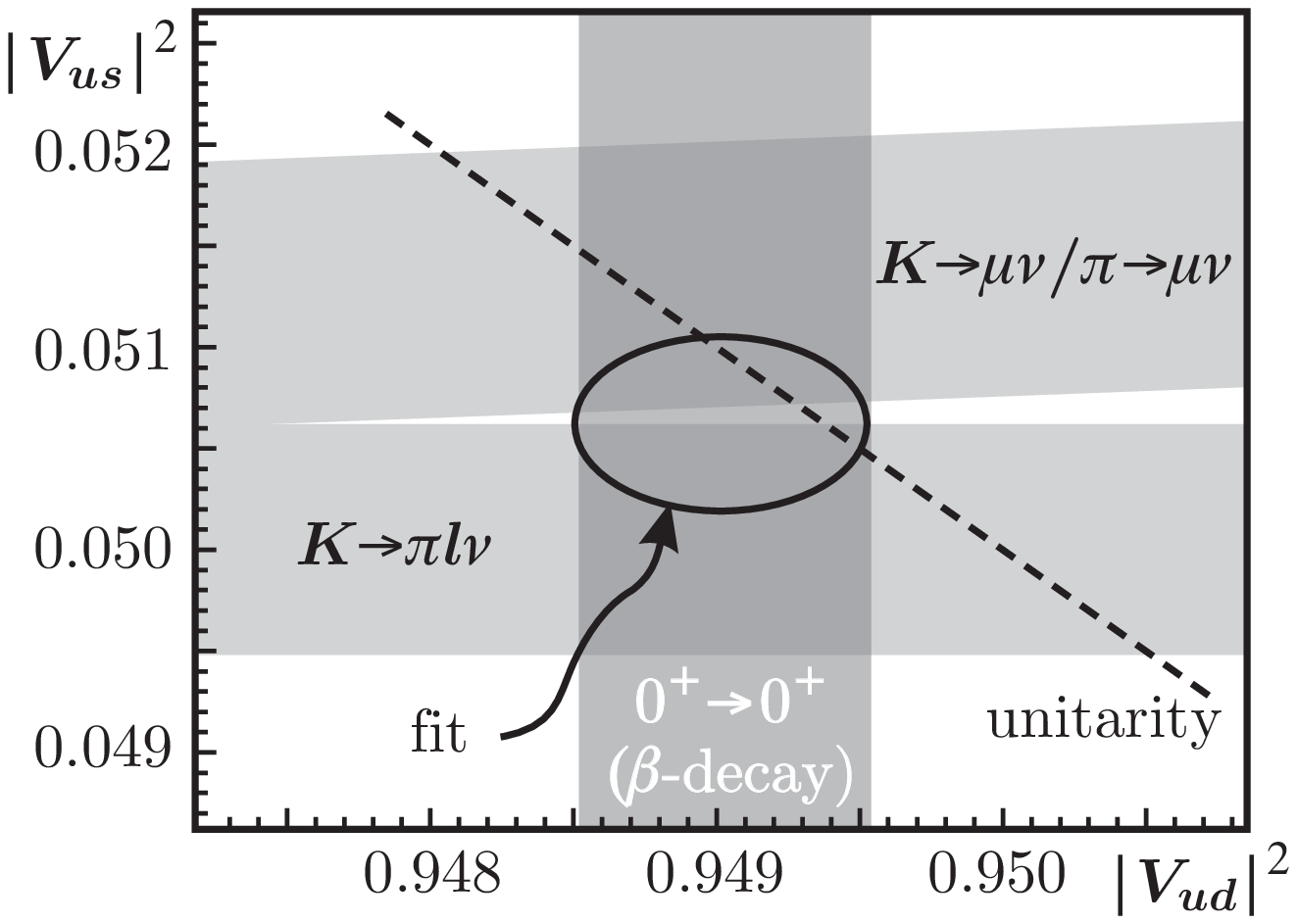}
\includegraphics[width=5.5cm]{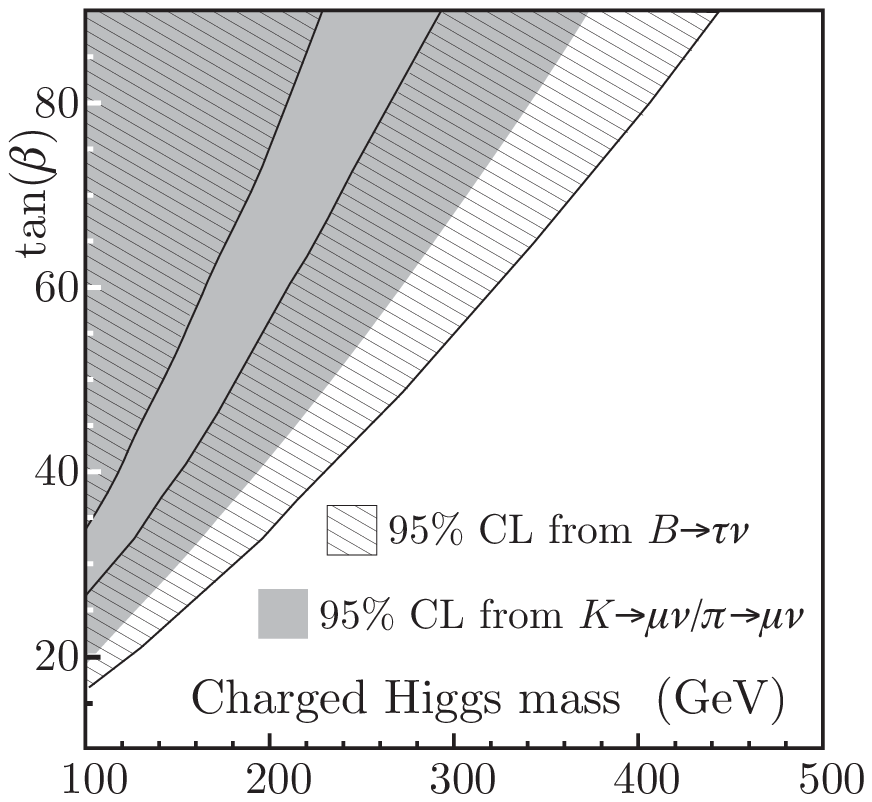}
\caption{\label{fig:fig1}Left: Results of fits to $|\vud|$, $|\vus|$, and
  $|\vus|/|\vud|$. Right: Excluded region in the $m_{H^+} - \tan\beta$ plane by the measurement
$R_{\ell23}$; the region excluded by $B \rightarrow \tau \nu$ is also indicated.}
\end{figure}

Defining the ratio $r_{\mu e}= |\fVus|_{\mu3}^2 /|\fVus|_{e3}^2$
and using eq.\ref{eq:gammakl3}, we have $r_{\mu e} = g_{\mu}^2/ g_{e}^2$,
with $g_{\ell}$ the coupling strength at the $W \to \ell \nu$ vertex.
Lepton universality can be then tested comparing the measured value of $r_{\mu e}$ 
and its Standard Model prediction $r_{\mu e}^{SM}=1$. Averaging between charged and neutral modes, 
we obtain $r_{\mu e} =1.000(8)$, to be compared with what obtained from leptonic pion decays,
$(r_{\mu e})_{\pi} =1.0042(33)$~\cite{unilep_pi}, and from leptonic $\tau$ decays 
$(r_{\mu e})_{\tau} =1.0005(41)$~\cite{unilep_tau}.
Using the determination of $|\vus|\fzero$ from \kltre\ decays and 
the preliminary result $\fzero = 0.964(5)$ from the UKQCD/RBC Collaboration~\cite{UKQCD}, 
we get $|\vus| = 0.2237(13)$. 
Furthermore a measurement of $\vus/\vud$ can be obtained from a comparison of the radiative inclusive decay 
rates of $\kpm \rightarrow \mudue (\gamma)$ and $\pipm \rightarrow \mudue (\gamma)$ combined 
with a lattice calculation of $f_K/f_{\pi}$~\cite{marcianovusvud}.
Using $\BR(\kpm \rightarrow \mudue) = 0.6366(17)$ from KLOE~\cite{kmu2}
 and the preliminary lattice result $f_K/f_\pi = 1.189(7)$ from the HP/UKQCD '07~\cite{HP/UKQCD},
we get $\vus/\vud = 0.2323(15)$. 
This value can be used in a fit together with the measurements of
\vus\ from \kltre\ decays and $\vud = 0.97418(26)$~\cite{vudyetold}, 
as shown in figure~\ref{fig:fig1} left. 

The result of this fit is $\vud = 0.97417(26)$ and $\vus = 0.2249(10)$, with 
$\chi^2/ndf =2.34/1$ (Prob$ = 13\%$), from which we get $1-\vud^2 +
\vus^2+\vub^2=4(7)\times10^{-4}$ compatible with unitarity at 0.6$\sigma$ level. 
Using these results, we then evaluate 
$G_{CKM}=G_F (\vud^2 + \vus^2+\vub^2)^{1/2}=(1.16614\pm0.00040)\times 10^{-5}$ GeV$^{-2}$ 
which is in perfect agreement with the measurement from the muon
lifetime $G_F=(1.166371\pm0.000006)\times 10^{-5}$ GeV$^{-2}$ and is 
competitive with the present accuracy of the measurements 
from tau-lepton decays and electroweak precision tests.
\section{BOUNDS ON NEW PHYSICS}
\label{sec:np}
A particularly interesting observable is the ratio of the \vus\ values
obtained from helicity-suppressed and helicity-allowed kaon modes:
$R_{\ell23} = \left| \vus(K_{\ell2})/\vus(K_{\ell3}) \times V_{ud}(0^+\to0^+)/V_{ud}(\pi_{\mu2}) \right|$,
which is equal to 1 in the SM.
The presence of a  scalar current due to a charged Higgs $H^+$ exchange is expected to lower the value of $R_{\ell23}$\cite{rl23}:
\begin{equation}
R_{\ell23} = \left| 1 - \frac{m^2_{K^+}}{m^2_{H^+}}\left( 1 -  \frac{m^2_{\pi^+}}{m^2_{K^+}}\right)
\frac{\tan^2\beta}{1+0.01 \tan\beta} \right|,
\label{eq:rl23}
\end{equation}
with $\tan \beta$ the ratio of the two Higgs vacuum expectation values in the MSSM.
To evaluate $R_{\ell23}$, we fit our experimental data on $K_{\mu2}$ and $K_{\ell3}$ decays, using as external
inputs the most recent lattice determinations of \fzero\ ~\cite{UKQCD} and $f_K/f_\pi$
~\cite{HP/UKQCD}, the value of \vud\ from \cite{vudyetold},
and $\vud^2 + \vus(K_{l3})^2 = 1$ as a constraint.
We obtain $R_{\ell23} = 1.008 \pm 0.008$~\cite{allvus}.
Figure \ref{fig:fig1} right shows the region excluded at 95\% CL in the
charged Higgs mass $m_{H^+}$ and $\tan\beta$ plane together with
the bounds from BR$(B \rightarrow \tau \nu)$~\cite{btaunu}.
The figure shows that Kaon physics and B physics provide complementary information.

\end{document}